# Electric-field control of hydrogen bonding via interfacial charge at atomic resolution

Nassar Doudin[1†*], Jian Jiang[2†], Chun Tang[3,4], Xiao Cheng Zeng[5,6*] and Mohammed Th. Hassan[1*].

[1]Department of Physics, University of Arizona, Tucson, AZ 85721, USA.
[2]Shenzhen Research Institute, City University of Hong Kong, Shenzhen, P. R. China.
[3]Faculty of Civil Engineering and Mechanics, Jiangsu University, Zhenjiang, Jiangsu, China.
[4]Suzhou Laboratory, Suzhou, China.
[5]Department of Material Science and Engineering, City University of Hongkong, Kowloon, Hongkong, China.
[6]Hong Kong Institute for Clean Energy, City University of Hong Kong, Kowloon, Hong Kong, China

† These authors contributed equally: Nassar Doudin, Jian Jiang
* Corresponding authors: ndoudin@arizona.edu, xzeng26@cityu.edu.hk and mohammedhassan@arizona.edu

Hydrogen-bond networks govern molecular structure and function across chemistry, biology and materials science, yet their deterministic control at the atomic scale remains a central challenge.[1–9] **Here, we directly visualize how an external electric field enables reversible control of a hydrogen-bond network in monolayer ice on graphite through interfacial charge redistribution. Low-temperature scanning tunnelling microscopy reveals a field-driven transition from a mobile, physisorbed, non-wetting water phase to an ordered hexagonal monolayer, enabling deterministic nucleation, growth and complete wetting on an otherwise inert surface. Systematic variation of the field induces continuous lattice strain coexisting with discrete conductance states, revealing coupled structural and electronic responses. Reversal of the field polarity drives collective dipolar inversion, enabling switching between symmetry-equivalent configurations without disrupting the lattice. Supported by first-principles theory and bias-dependent imaging, these effects arise from field-induced modification of the interfacial electronic structure rather than purely geometric or orientational effects. These results establish interfacial charge redistribution as a general mechanism for electrically programming hydrogen-bond networks, providing a route to control molecular organization, electronic properties and collective dipolar order at interfaces.**

**Introduction:**

Hydrogen bonding is a fundamental interaction that governs the structure, dynamics and function of molecular systems, from biological assemblies to molecular materials.[9–14] It underlies processes ranging from molecular recognition and enzymatic activity to supramolecular assembly and interfacial phenomena.[13,14] Beyond defining molecular organization, hydrogen bonds can also mediate charge transfer, enabling long-range electronic coupling and tunable transfer in molecular systems.[12,15–17] This dual structural and electronic character suggests that hydrogen-bond networks can, in principle, be controlled through electronic degrees of freedom.[16,18] However, achieving deterministic control at the atomic scale remains a major challenge, particularly at interfaces where configurational degeneracy, thermal fluctuations and weak molecule–substrate interactions dominate. [6,19–22]

Water serves as a stringent and broadly relevant model system for addressing this fine molecular control challenge.[2,23–25] As the prototypical hydrogen-bonded molecular system, its structure and dynamics are governed by a highly cooperative, fluctuating network that is exquisitely sensitive to external electrical perturbations.[2,23,26–29] At interfaces, this sensitivity is amplified: on weakly interacting substrates such as graphite, water typically forms a highly mobile phase that resists nucleation into ordered structures.[5,7,30,31] This behaviour reflects a delicate balance between intermolecular hydrogen bonding and molecule–substrate interactions that is difficult to control deterministically.[5,32–34] Electric fields provide a direct and local means to perturb this balance by coupling to molecular dipoles and modifying the interfacial electronic environment.[3,35–37] Such field–matter interactions are central to processes ranging from electrochemistry to catalysis and biological function,[38–41] yet their microscopic impact on hydrogen-bond stability remains inadequately understood. In particular, a quantitative framework linking electric field, interfacial charge redistribution and hydrogen-bond stability at atomic resolution is lacking.

In this work, we demonstrate that an external electric field enables direct and reversible control of a hydrogen-bond network in monolayer water on graphite through interfacial charge redistribution. Using low-temperature scanning tunnelling microscopy, we directly visualize a field-driven transition from a mobile, physisorbed, and non-wetting water phase to an ordered hexagonal monolayer, establishing deterministic nucleation, growth, and complete wetting on an otherwise hydrophobic surface. By tuning the electric field, we access regimes spanning stable ordering, dynamic reconfiguration and field-induced instability, accompanied by continuous lattice strain, discrete conductance states, and polarity-driven dipolar switching. These observations reveal that hydrogen-bond stability is governed by field-induced modification of the interfacial electronic structure rather than purely geometric or orientational effects, establishing a general mechanism for electrically programming hydrogen-bond molecular networks at interfaces, enabling control of molecular organization and function in aqueous environments.

**Results:**

Graphite provides a structurally well-defined and hydrophobic platform for probing interfacial water and ice. The mechanism and step-by-step field-induced nucleation of complete monolayer ice layer on graphite is shown in Fig. 1. Large-area scanning tunnelling microscopy (STM) images show an atomically clean surface with no detectable contamination (Fig. 1a). Atomic-resolution imaging (inset, Fig. 1a) resolves the characteristic hexagonal lattice of graphite, confirming structural integrity and providing a reference for subsequent measurements.

STM images acquired after exposure to a small dose of water at 76 K (Fig. 1b) show diffuse, poorly defined features (cyan arrows). The absence of stable atomic contrast in the high-resolution inset indicates that the adsorbed water molecules remain highly mobile, diffusing on timescales faster than STM imaging. Hence, STM images acquired after increasing the water dose (Fig. 1c) show the formation of an extended two-dimensional mobile phase, in which intermolecular interactions are insufficient to stabilize a persistent hydrogen-bond network. The inset shows a magnified view of a single terrace, displaying diffuse and streak-like features (highlighted by arrows) characteristic of this mobile phase. These observations indicate that, in the absence of external control, interfacial water remains dynamically disordered, even at temperature as low as 76 K, and does not nucleate into an ordered phase. Reducing the bias induces a qualitative transition by modifying the local electric field in the STM junction. Under these conditions, the mobile water phase condenses abruptly into ordered domains (Fig. 1d), forming a partially complete monolayer (~0.8 ML) with a hexagonal arrangement coexisting with bare graphite (highlighted by violet arrow). The inset (Fig. 1d, top terrace) resolves the ordered ice structure at atomic resolution. This abrupt emergence of long-range order indicates that the electric field is the key parameter governing hydrogen-bond network stabilization and reducing the nucleation barrier for the transition from mobile to ordered water phase. The apparent height of the hexagonal layer relative to the graphite surface is measured to be $3.3 \pm 0.1$ Å, using the sub-monolayer regime (Fig. S1) as a reference, which is in a good agreement with first-principles calculations, as discussed below. Further water exposure approaching a fully saturated first layer under the same conditions leads to the formation of a continuous two-dimensional ice overlayer (Fig. 1e), extending seamlessly across terraces and indicating complete wetting of the graphite surface. The inset of Fig. 1e highlights the seamless extension of the ice layer across adjacent terraces. A magnified STM image (Fig. 1f) shows a single structural phase extending across terraces with long-range hexagonal order, as confirmed by Fourier analysis (right inset). The atomic-resolution image (left inset) resolves the hexagonal lattice of the ice overlayer with a periodicity of $5.00 \pm 0.05$ Å. Additional atomically resolved STM images (Fig. S2) reveal a defect-free monolayer with no domain boundaries or structural distortions, indicating uniform nucleation and stabilization of a single thermodynamic phase under the applied field.

These observations demonstrate that monolayer ice nucleation on graphite occurs under an applied electric field, in contrast to the previously reported bilayer interlocked hexagonal ice at zero field.[42] The long-range order of monolayer hexagonal ice on graphite—otherwise suppressed by weak molecule–substrate interactions—is deterministically activated by an external electric field. This arises from field-induced modification of the interfacial electronic environment, which stabilizes the hydrogen-bond network through charge redistribution and dipolar alignment. A schematic (Fig. 1g) illustrates the coupling of the electric field to molecular dipoles and interfacial charge, while structural models (Fig. 1h,i) support a hexagonal monolayer ice network with each water molecule forming three hydrogen bonds and one dangling bond that aligns with the applied vertical field. Together, these results demonstrate that the electric field drives a transition from a dynamically disordered, non-wetting water layer to a uniformly ordered hydrogen-bond network, enabling deterministic nucleation and growth of interfacial ice at the atomic scale.

Having established that a uniform monolayer ice film can be deterministically nucleated under electric-field control, we next determine how the hydrogen-bond network evolves under increasing electric field strength. By imaging the same region while systematically tuning the voltage bias (from +0.6 to +1.75V), we directly resolve how the applied field reshapes the energy landscape of the ice monolayer at molecular resolution. At mild bias (~+0.85 V), the ice layer exhibits a well-ordered hexagonal structure with stable, time-invariant contrast (Fig. 2a), indicating that the hydrogen-bond network occupies a single, well-defined free-energy minimum. In this regime, the structure is robust against perturbations, consistent with a stable equilibrium configuration. As the bias is increased, the system enters a qualitatively distinct regime characterized by dynamic reconfiguration. At ~+1.35 V, molecular-scale contrast fluctuations emerge (Fig. 2b), reflecting reversible switching between two distinct hydrogen-bond configurations, likely corresponding to different molecular dipole orientations and associated local electronic structures. These bistable features indicate that the applied electric field reduces the energy barrier separating competing hydrogen-bond arrangements, enabling access to multiple local minima in the free-energy landscape. The ice monolayer thus becomes dynamically reconfigurable, with its configuration continuously tunable by the field. Further increasing the bias drives the system beyond this bistable regime. Spatially resolved imaging across regions scanned at different biases (Fig. 2c) reveals a progressive loss of order, culminating in local disruption of the hydrogen-bond network at the highest voltages, with the onset of melting and desorption observed at $V_s \sim +1.65$ V. Notably, within a single scan, regions acquired at different biases display abrupt transitions in appearance with well-defined boundaries (cyan arrow, Fig. 2c), ruling out changes in topography or tip condition and directly reflecting bias-dependent variations in the local electronic structure. In this regime, strong electric fields and localized energy injection destabilize intermolecular interactions, driving the system into a non-equilibrium state in which long-range order cannot be sustained.

Importantly, these field-induced transitions are highly reproducible. Repeated measurements under controlled bias conditions consistently yield the same sequence of structural transformations (Fig. 2d,e), demonstrating that the ice layer can be deterministically cycled between stable, bistable and unstable regimes. Notably, analogous behaviour is observed at negative bias (Fig. S3), indicating that the structural response is governed by the electric field rather than the polarity of the applied voltage. This reproducibility establishes the electric field as a continuous and reliable control parameter governing interfacial water at the atomic scale.

Comparison with simulated STM images (Fig. 2f), supported by empty-state STM measurements (Fig. S2b and the upper region of Fig. 2e), reproduces the site-dependent contrast associated with specific molecular configurations (blue arrow), confirming that variations in tunnelling current reflect changes in the local electronic structure. This agreement shows that the applied electric field simultaneously modulates both the geometry and electronic structure of the hydrogen-bond network.

To gain deeper insight into the field-induced effects on the monolayer ice, extensive classical molecular dynamics (MD) simulations were performed. The average height of the monolayer ice above the substrate increases with applied electric field (Fig. 2g), reflecting puckering of the ice layer that reduces the field–water interaction energy. At zero field, the computed height is approximately 3.3 Å, consistent with experimental measurements and confirming the formation of a true monolayer rather than a bilayer. Component analysis of the interaction energies reveals a dramatic decrease in the water–field interaction (Fig. 2h), which drives alignment of the water molecular dipoles along the field direction and induces puckering of the monolayer. In contrast, the applied electric field weakens both the water–water and water–substrate interactions, demonstrating that the stability of the monolayer ice on this hydrophobic surface is primarily field-induced. The melting temperature of the monolayer ice increases under moderate fields (below ~1 V Å$^{-1}$), indicating field-enhanced stability, whereas at higher field strengths the hydrogen-bond network weakens, rendering the monolayer unstable (Fig. 2i).

The results presented in Figure 2 establish that an external electric field continuously reshapes the energy landscape of the hydrogen-bond network, enabling deterministic access to equilibrium, bistable and non-equilibrium states within a single molecular layer. Next, we establish a direct and quantitative coupling between the structural and electronic properties of the monolayer ice film under an applied electric field. By probing the same region while systematically varying the bias, we resolve how the hydrogen-bond network responds as a coupled electromechanical system, in which structural distortions and electronic transport are intrinsically linked.

As shown in Fig. 3, as the electric field increases, the ice lattice undergoes a continuous and reproducible deformation. High-resolution STM images reveal a systematic change in intermolecular spacing (Fig. 3a–c), establishing that the hydrogen-bond network experiences field-induced lattice strain. The extracted lattice constant of the hexagonal monolayer ice decreases monotonically with increasing applied electric field (Fig. 3d), demonstrating that intermolecular interactions are continuously tuned by the interfacial electrostatic environment. The reduced lattice constant reflects both in-plane shrinkage and out-of-plane puckering of the hexagonal rings, consistent with first-principles calculations that reproduce the same monotonic trend. The ice monolayer thus behaves as a mechanically compliant lattice whose structure is directly controlled by the applied field.

In striking contrast to the continuous structural response, the electronic behaviour is intrinsically discrete. STM images acquired at different bias voltages (Fig. 3e–g), under constant-current conditions, show abrupt changes in contrast, indicating that the system accesses distinct electronic states stabilized by specific hydrogen-bond configurations. These states appear as well-defined tunnelling regimes with reproducible contrast at a given bias. Atomically sharp boundaries between regions of different contrast within a single scan (Fig. 3g,h) demonstrate spatial coexistence of these states and exclude gradual variations in imaging conditions, indicating switching between metastable configurations separated by finite energy barriers.

Spatially resolved measurements further reveal the threshold nature of these transitions. At low bias (~0.3 V), the tunnelling current is strongly suppressed, consistent with the intrinsic electronic gap of the ice layer. Above a threshold (~0.53 V), conduction is abruptly activated, marking a transition to a higher-conductance regime. The reproducibility, spatial locality and abruptness of these transitions (Fig. 3e–i) establish that they arise from intrinsic changes in the electronic structure rather than experimental perturbations.

First-principles calculations show that the applied electric field drives a continuous evolution of the electronic structure, including a transition from insulating to semiconducting and ultimately metallic behaviour (see below), consistent with continuous electronic tuning underlying discrete, state-dependent tunnelling responses. Independent non-contact AFM measurements (Fig. 3j) resolve a well-defined hexagonal ice network on the graphite surface, reproducing the lattice constant measured by STM. The observed contrast is consistent with short-range repulsive interactions between the tip and upward-oriented hydrogen atoms, as indicated by the overlaid structural model, in agreement with theoretical predictions (Fig. 1h,i) and empty-state STM imaging (Fig. S2b). These results directly link the electronic transitions to field-induced modifications of the hydrogen-bond network.

Water and ice are well-known wide-bandgap insulators. Hybrid DFT calculations at the HSE06 level yield a band gap of approximately 6 eV for the monolayer ice at zero electric field (Fig. 3k), resulting in negligible charge-density overlap with the underlying graphene. In contrast, application of a high electric field (1 V Å$^{-1}$) strongly polarizes the system along the field direction and induces pronounced interfacial charge redistribution (Fig. 3i). This is accompanied by a substantial reduction in the band gap, driving an insulator-to-metal transition while the atomic structure remains largely unchanged. Both GGA (PBE) and hybrid-GGA (HSE06) calculations show that the band gap of the monolayer ice decreases with increasing electric field up to ~0.8 V Å$^{-1}$ and then rises slightly to approximately 1 eV at higher fields, owing to structural changes associated with puckering of the monolayer ice (Fig. 3m).

The results in Figure 3 reveal a physical principle: an external electric field drives continuous lattice strain in a hydrogen-bond network that is transduced into discrete electronic states, providing a direct mechanism to control charge transport through molecular structure at the atomic scale.

Furthermore, we demonstrate that electric-field polarity enables reversible switching between symmetry-equivalent structural states in a monolayer ice film, revealing a collective dipolar degree of freedom intrinsic to the hydrogen-bond network.

STM imaging under opposite bias polarities reveals a pronounced and reproducible inversion of contrast across the hexagonal lattice. Under positive bias, the ice monolayer exhibits a characteristic arrangement of bright and dim features as presented in Fig. 4a. Reversing the polarity produces a complementary contrast pattern over the same region as can be seen in Fig. 4b, while preserving lattice symmetry, periodicity and long-range order. This invariance of the lattice, combined with the inversion of electronic contrast, indicating that the transformation arises from an internal reconfiguration of the hydrogen-bond network rather than structural rearrangement or imaging artefacts.

To identify the microscopic origin of this transition, we compare the experimental observations with optimized structural models (Fig. 4c,d). The calculations reveal two nearly degenerate configurations corresponding to opposite molecular dipole orientations (H-up and H-down). Charge density difference analysis (Fig. S5) reveals distinct interfacial charge redistribution between the two configurations, which modulates the local electronic structure and accounts for the observed contrast inversion in STM.

The switching is driven by direct coupling between the applied electric field and the intrinsic dipole moments of water molecules. The field stabilizes one dipolar configuration, and reversal of its polarity drives a collective inversion of dipole orientation across the lattice, thereby switching the hydrogen-bond network between symmetry-equivalent, electronically distinct states without perturbing the underlying structure. A schematic representation (Fig. 4e,f) illustrates this mechanism, showing the field-driven alignment of molecular dipoles and the resulting reconfiguration of the hydrogen-bond network.

Crucially, the transition is fully reversible and reproducible, demonstrating deterministic switching between two states under repeated polarity cycling. The ice monolayer thus realizes a bistable dipolar system at the atomic scale, controlled solely by the direction of the electric field. Fig. 4 establishes a general mechanism: electric-field polarity couples to molecular dipoles in a hydrogen-bond network, enabling reversible switching between symmetry-equivalent structural states and providing a direct route to control collective molecular order at interfaces.

**Discussion:**

The results establish interfacial charge redistribution as a direct mechanism for controlling hydrogen-bond networks at the atomic scale. By coupling an external electric field to the intrinsic dipole moments of water molecules, the system can be driven between distinct structural and electronic states in a deterministic and reversible manner. Crucially, this control extends beyond simple dipolar alignment: the combined experimental and theoretical results demonstrate that the applied field modifies the interfacial electronic structure, which in turn controls hydrogen-bond stability, lattice structure and electronic transport.[3,38,39,43–45]

A key finding is the coupled electromechanical response of the hydrogen-bond network. The applied field induces a continuous lattice strain, reflecting the mechanical compliance of the intermolecular framework, while the electronic response is intrinsically discrete, with well-defined conductance states separated by finite energy barriers.[39,46–48] This coexistence of continuous structural tuning and discrete electronic transitions demonstrates a direct transduction of molecular-scale distortions into electronic functionality,[49] highlighting hydrogen-bond networks as responsive molecular systems in which structural and electronic degrees of freedom are intrinsically linked.

The polarity-dependent switching reveals a collective dipolar degree of freedom analogous to ferroelectric behaviour,[50,51] realized within a hydrogen-bonded molecular system.[52] The ability to reversibly switch between symmetry-equivalent configurations without perturbing the underlying lattice highlights the cooperative nature of the network and its sensitivity to the interfacial electrostatic environment. Importantly, this behaviour arises from field-induced modulation of the electronic structure rather than purely geometric rearrangement,[53] demonstrating that hydrogen-bond networks can support electrically addressable states with well-defined structural and electronic signatures.

These findings establish a general framework for controlling molecular organization at interfaces. Hydrogen-bond networks are central to aqueous systems, biological function and soft materials,[54,55] and their electrical programmability suggests new routes for tuning reactivity, transport and intermolecular interactions. More broadly, electric-field coupling to molecular dipoles, mediated by interfacial charge, offers a strategy for designing responsive molecular systems in which structural, electronic and dipolar degrees of freedom are intrinsically linked. Together, this work establishes a direct route to deterministically program hydrogen-bond networks and control molecular order and function at interfaces.

**Conclusion:**

Electric-field-induced interfacial charge redistribution provides a physically grounded mechanism for deterministic and reversible control of hydrogen-bond networks at the atomic scale. In monolayer ice on graphite, this coupling enables controlled nucleation, continuous structural tuning and discrete conductance switching, together with polarity-driven dipolar inversion within a unified framework. These results demonstrate that hydrogen-bond networks can be electrically programmed through their interfacial electronic structure, establishing a general principle for controlling molecular organization and function at interfaces, with potential implications for probing biomolecular structure and dynamics at interfaces in controlled aqueous environments.[56]

**Methods**

**1- STM and AFM experiments**

All experiments were performed using an ultra-high vacuum Scienta Omicron POLAR STM/AFM system operated at 76 K. Electrochemically etched tungsten (W) tips were used for STM, and a qPlus sensor equipped with a W tip (resonance frequency $f_0 = 24.7$ kHz, spring constant $k_0 \approx 1.8$ kN m$^{-1}$, quality factor $Q \approx 43,000$) was used for AFM measurements.

Highly Oriented Pyrolytic Graphite substrates were cleaved ex situ in air using adhesive tape, producing atomically flat terraces extending over several hundred nanometres. The samples were immediately transferred into the ultra-high vacuum chamber (base pressure $< 1 \times 10^{-10}$ mbar) and subsequently annealed to 800 K prior to water exposure. Surface order and cleanliness were verified by STM.

Ultrapure water (Sigma-Aldrich, 99.9%) was further purified by three to five freeze–pump–thaw cycles under vacuum to remove residual gas impurities. Water molecules were dosed at 76 K through a directional tube doser positioned approximately 10 cm from the sample.

The sample temperature was monitored using a DT-670 silicon diode (Lake Shore Cryotronics). STM images were acquired in constant-current mode at 76 K, with bias voltages referenced to the sample (scan parameters are provided in the figure captions). The observed dynamics and structural changes were verified to be independent of STM tip effects and tip instabilities. Control experiments across multiple surface regions and a wide range of tunneling parameters yielded consistent and reproducible results, indicating that the observed behavior reflects genuine surface processes.

All frequency-shift (Δ$f$) AFM images were acquired in constant-height mode with an oscillation amplitude of 1 Å using a CO-terminated tip. The CO-functionalized tip was prepared by positioning the tip over a CO molecule adsorbed on Au(111) surface and applying a bias of approximately 2.5 V. For monolayer ice on HOPG, AFM tip heights are referenced to the STM set point (0.50 V, 100 pA) on the HOPG surface and the first ice layer, respectively. All images were processed using WSxM.[57]

2- **STM simulation**

STM images were simulated applying the Tersoff–Hamann approach[58] at a constant height of about 2 Å above the ice surface. The tunnelling current was calculated by integrating the local density of states (LDOS) over the energy window corresponding to the applied bias. Along the $z$ axis, a minimum empty space of 15 Å was chosen to avoid interactions between replicated cells. A dipole correction[59] was applied to cancel spurious Columbic interactions among replicated images in the case of asymmetric unit cells.

3- **First-principles calculations**

Density functional theory calculations were performed using the Vienna ab initio Simulation Package (VASP 6.4).[60,61] The exchange–correlation energy was treated at the level of the Perdew–Burke–Ernzerhof (PBE)[62] generalized gradient approximation and the screened hybrid HSE06 functional.[63] A plane-wave kinetic energy cutoff of 550 eV was employed. The electronic self-consistent field (SCF) convergence criterion was set to 1 × 10$^{-6}$ eV, and the ionic relaxation convergence criterion was 0.01 eV Å$^{-1}$. An external electric field was applied perpendicular to the monolayer ice plane. Dipole corrections were included to eliminate spurious electrostatic interactions between periodic images. A vacuum spacing of 50 Å along the out-of-plane direction was used to minimize field-induced finite-size effects on the electronic band structure.

**4- Classical MD simulations**

Classical molecular dynamics (MD) simulations were performed using the Large-scale Atomic/Molecular Massively Parallel Simulator (LAMMPS).[64] The TIP4P/Ice[65] water model was employed for water molecules. The oxygen–carbon interaction was described by the Lennard-Jones potential with parameters σ = 3.191 Å and ε = 0.0046 eV. A timestep of 0.5 fs was used. The melting point of the monolayer ice was calculated by heating the system at a constant rate of 10 K ns$^{-1}$ for a total of 30 ns. The melting temperature was determined as the point at which the structure of the monolayer ice changed upon increasing temperature.

**Acknowledgments**

This project is funded by Gordon and Betty Moore Foundation Grant GBMF#11476 to M. Th. H (https://www.moore.org/grant-detail?grantId=GBMF11476). This material is also based upon work partially supported by the Air Force Office of Scientific Research under award number FA9550-22-1-0494 to M. Th. H. X.C.Z. acknowledges support from the Hong Kong Global STEM Professorship Scheme and the Research Grants Council of Hong Kong (GRF Grant No. 11204123 and No. 11302225). J.J. acknowledges the funding support of the National Natural Science Foundation of China (Grant No. 22303072). We thank Carson Bayer, Mohamed Sennary and Aalekhya Saha (University of Arizona) for their assistances.

**Author contributions:**

N.D. conceived, designed, performed the experiments, analysed the data, and carried out the STM simulations, and wrote the original draft of the manuscript. J.J. and C.T. carried out the simulations and calculations under the supervision of X.C.T. M. Th.H. conceived, evaluated analysed, interpreted the experimental data and supervised the project; All authors discussed the results and contributed to the final manuscript.

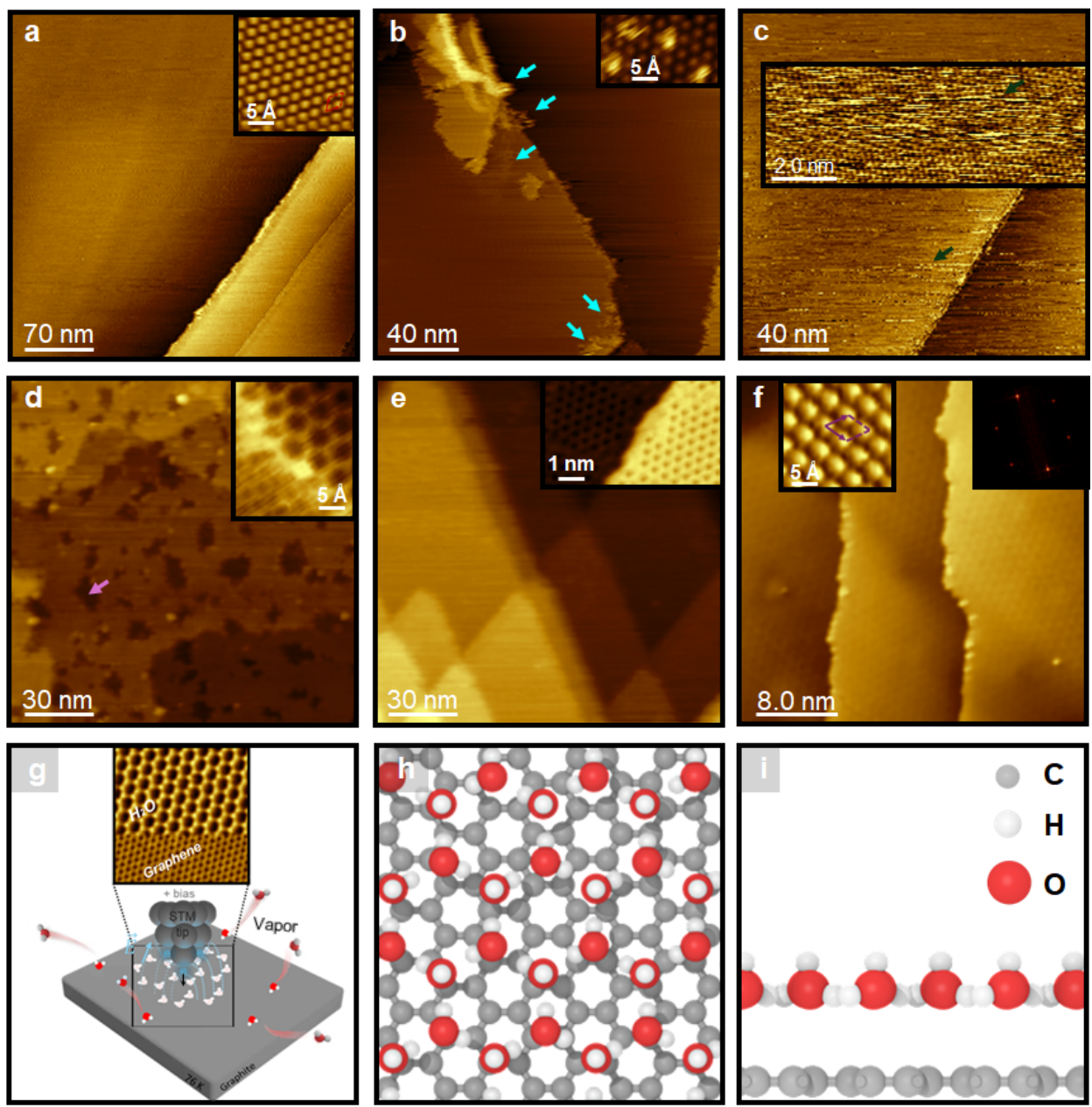

**Fig. 1 | Electric-field control of hydrogen-bond ordering in interfacial water and ice on graphite.** a. Large-area STM image of pristine graphite (350 × 350 nm², Voltage Bias, $V_S$ = +1.9 V, tunneling current, $I_t$ = 250 pA), showing a perfectly clean surface. Inset (2.5 × 2.5 nm², $I_t$ = +0.21 V, $I_t$ = 250 pA): atomic-resolution image resolving the hexagonal lattice (lattice constant ≈ 2.46 Å, red dashed line). **b–f**, STM image sequence of the formation of a first ice layer recorded at 76 K. **b,** STM image (200 × 200 nm², $V_S$ = +1.3 V, $I_t$ = 250 pA) acquired after exposure to 0.05 L of water at 76 K (L, 1 L = 1.33 × 10⁻⁶ mbar·s), showing fuzzy features (cyan arrows) indicative of highly mobile water molecules. Inset: high-

resolution image (2.5 × 1.5 nm², $V_S$ = +1.3 V, $I_t$ = 250 pA) showing blurred contrast or diffuse features associated with rapid molecular motion. **c,** STM image (200 × 200 nm², $V_S$ = +1.3 V, $I_t$ = 250 pA) acquired after increasing the water dose to 5 L (dosed at 5 × 10⁻⁹ mbar), showing an extended two-dimensional mobile phase in which no stable hydrogen-bond network is formed. Inset: atomic-resolution image (20 × 8 nm², $V_S$ = +1.3 V, $I_t$ = 250 pA) showing diffuse contrast or streak features (highlighted by green arrows) characteristic of molecular mobility. **d,** STM image (150 × 150 nm², $V_S$ = +0.75 V, $I_t$ = 250 pA) acquired at reduced bias, showing condensation of the mobile phase into ordered domains. A partially complete monolayer (~0.8 ML) with a hexagonal structure coexists with clean graphite. Inset: atomic-resolution image (2.5 × 3 nm², $V_S$ = +0.75 V, $I_t$ = 250 pA) resolving the ordered phase and graphite. **e,** STM image (150 × 150 nm², $V_S$ = +0.75 V, $I_t$ = 250 pA) acquired after further water exposure (2L, at 5 × 10⁻⁹ mbar) reveals a continuous two-dimensional hexagonal ice layer (see inset:(5 × 3 nm², $V_S$ = +0.75 V, $I_t$ = 250 pA) extending across terraces, indicative of complete wetting. **f,** Resolution STM image (40 × 40 nm², $V_S$ = +0.80 V, $I_t$ = 250 pA) showing a single structural phase extending across terraces. Left inset: atomic-resolution image (2.5 × 2.5 nm², $V_S$ = +0.80 V, $I_t$ = 250 pA) resolving the hexagonal unit cell (Plum line). Right inset: corresponding fast Fourier transform confirming long-range order. **g,** Schematic illustrating the role of the electric field ($\vec{E}$) and tunnelling electrons (current flow is symbolized by black arrow) in stabilizing the hydrogen-bond network at the graphite interface; field lines are shown in cyan. **h,i,** Top and side views of the monolayer ice structure on graphene, showing a hexagonal hydrogen-bond network. Grey, red and white spheres represent carbon, oxygen and hydrogen atoms, respectively.

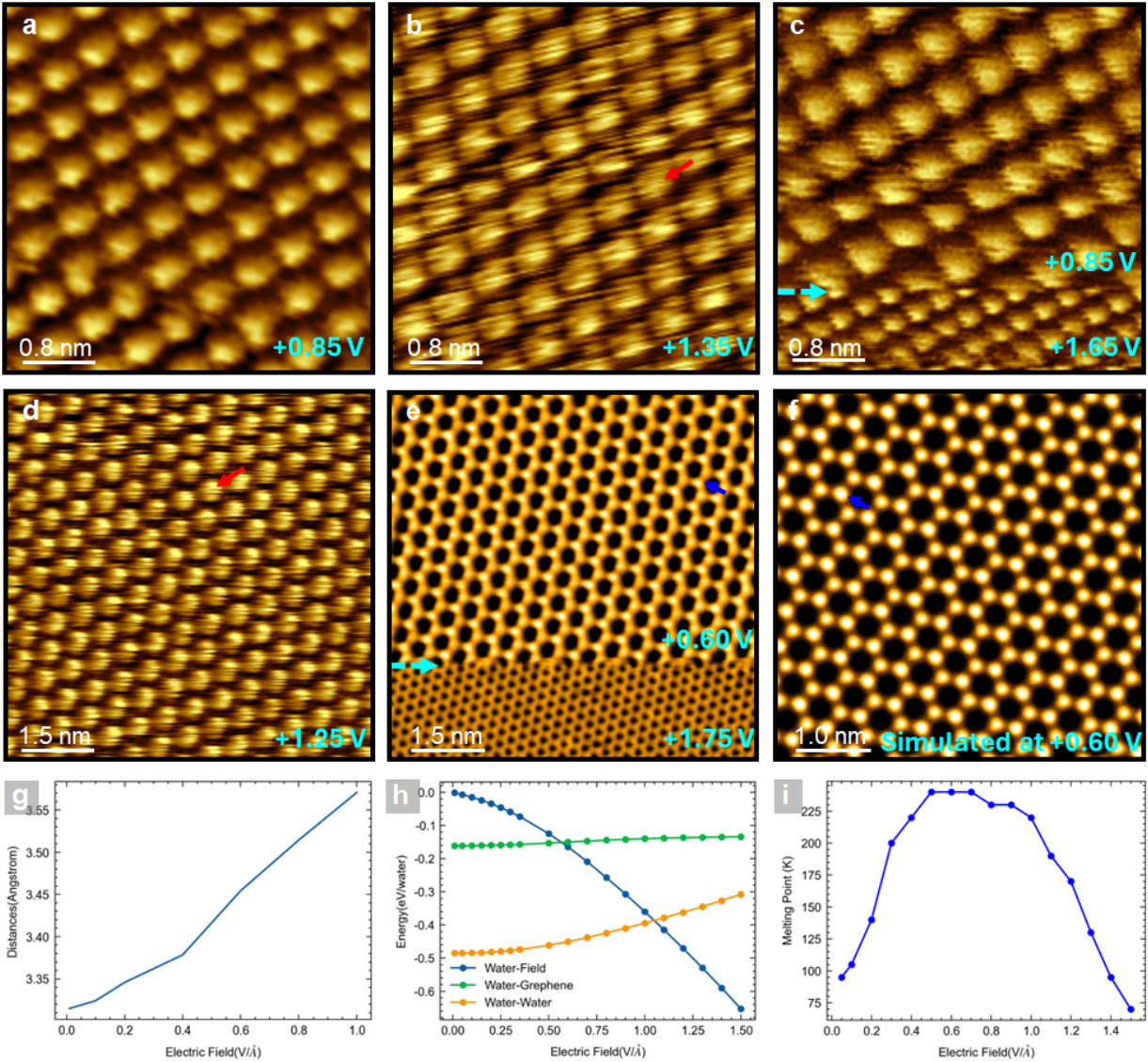

**Fig. 2 | Bias-dependent, electric-field-driven evolution of the hydrogen-bond network in interfacial ice on graphite at 76 K. a,** STM image (4 × 4 nm², $V_S$ = +0.85 V, $I_t$ = 280 pA) showing a well-ordered hexagonal ice structure with stable contrast. Scan direction: bottom to top. **b,** Increasing the bias (4 × 4 nm², $V_S$ = +1.35 V, $I_t$ = 280 pA) induces molecular-scale fluctuations, revealing reversible switching between two molecular configurations (red arrow). **c,** STM image (4 × 4 nm², $I_t$ = 280 pA) acquired across regions scanned at different bias voltages. The dashed cyan line marks the boundary between areas imaged at different biases; increasing bias progressively destabilizes the ice structure, with local disruption observed at $V_S ≈$ +1.65 V. **d&e,** STM images (7.5 × 7.5 nm², $I_t$ = 280 pA) demonstrating the reproducibility of bias-induced structural modifications. The same sequence of configurations is recovered upon repeated bias variation. **f,** Simulated STM image at $V_s$ = +0.60 V showing site-dependent contrast associated with the H-up configuration; the blue arrow marks a high-contrast site arising from variations in the local electronic structure. **g,** Average height of the monolayer ice from the substrate at different electric fields, obtained from molecular dynamics simulations. **h,** Component

analysis of the interaction energy from MD simulation. i, Calculated melting point of the monolayer ice at different electric fields.

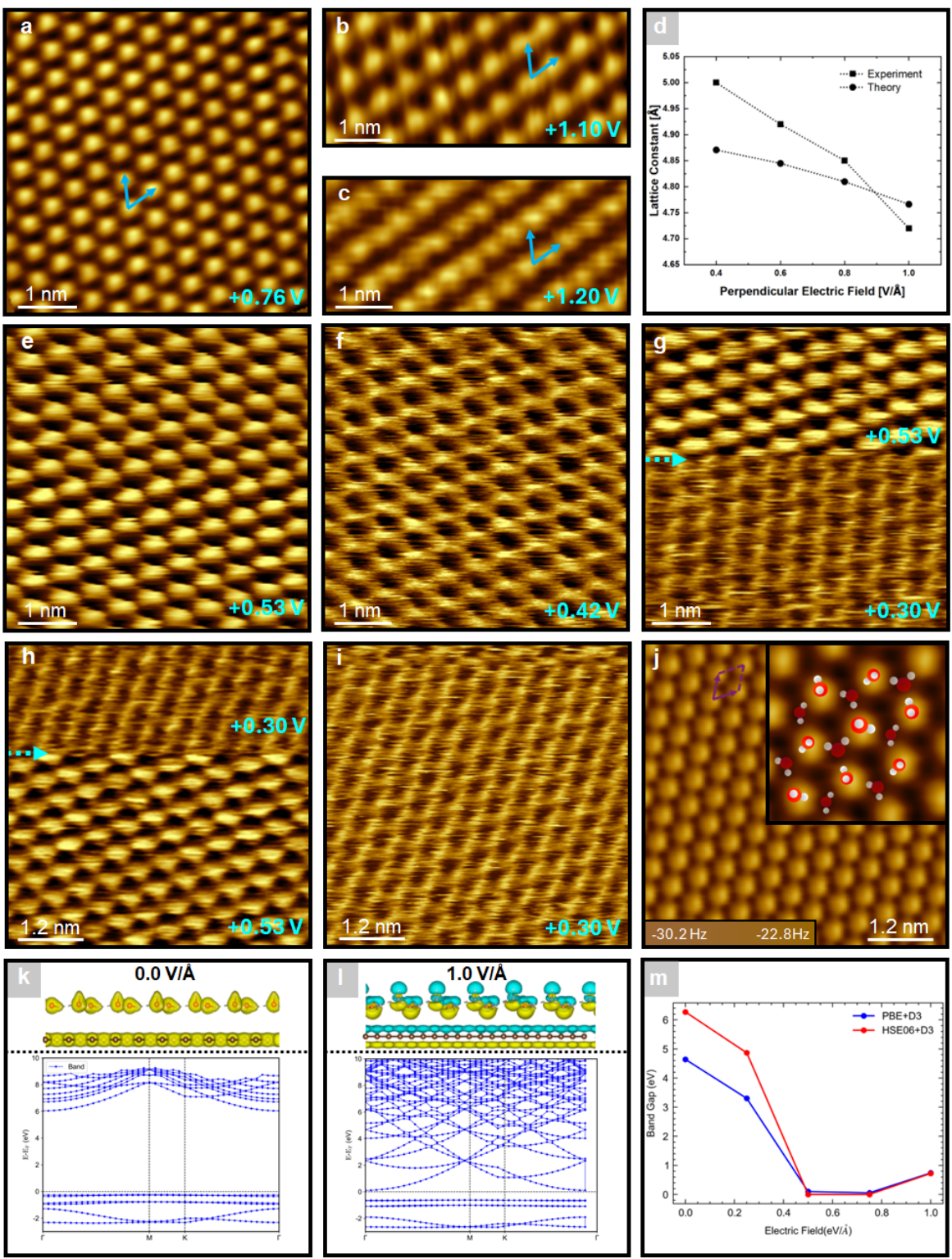

**Fig. 3 | Bias-dependent (Field controlled) lattice strain and quantized conductance in monolayer ice. a-c,** STM images of the same region acquired at increasing bias voltages, showing a systematic change in intermolecular spacing within the ice lattice (light blue line) (**a,** 5 × 5 nm², $V_S$ = +0.76 V; **b,** 5 × 2.1 nm², $V_S$ = +1.1 $V_S$; **c,** 5 × 2.1 nm², $V_s$ = +1.2 V; all $I_t$ = 300 pA). **d,** Lattice constant as a function of the electric field applied perpendicular to the surface. **e–g,** STM images (5 × 5 nm², $I_t$ = 300 pA) acquired at different bias voltages ($V_S$ = +0.53, +0.42 and +0.35 V), showing three distinct tunnelling regimes characterized by high, intermediate and strongly suppressed contrast, respectively. All images were recorded under constant-current conditions, such that variations in contrast reflect changes in the local electronic structure and tunnelling probability The transitions between these regimes are abrupt rather than continuous, indicating switching between discrete electronic states of the hydrogen-bond network. In g, an atomically sharp boundary (cyan arrow) separates regions of distinct contrast within a single scan, demonstrating spatial coexistence of these states and excluding gradual variations in imaging conditions. **h,i**, STM images demonstrating the reproducibility and spatial coexistence of multiple conductance states within a single scan. The cyan arrow marks the boundary between regions of distinct conductance. **j,** Non-contact AFM image of the ice monolayer on graphite acquired at 76 K (oscillation amplitude 120 pm, zero bias), resolving the hexagonal lattice. The plum line indicates the lattice periodicity; inset, structural model overlaid on the high-resolution image. **K,** Charge distribution obtained from DFT calculations (upper panel) and the band structure of monolayer ice in the absence of an electric field (lower panel). **l,** Charge difference distribution relative to the zero-field system obtained from DFT calculations (upper panel) and the band structure of monolayer ice at an electric field of 1 V Å$^{-1}$ (lower panel). **m,** Calculated band gap of monolayer ice as a function of electric field.

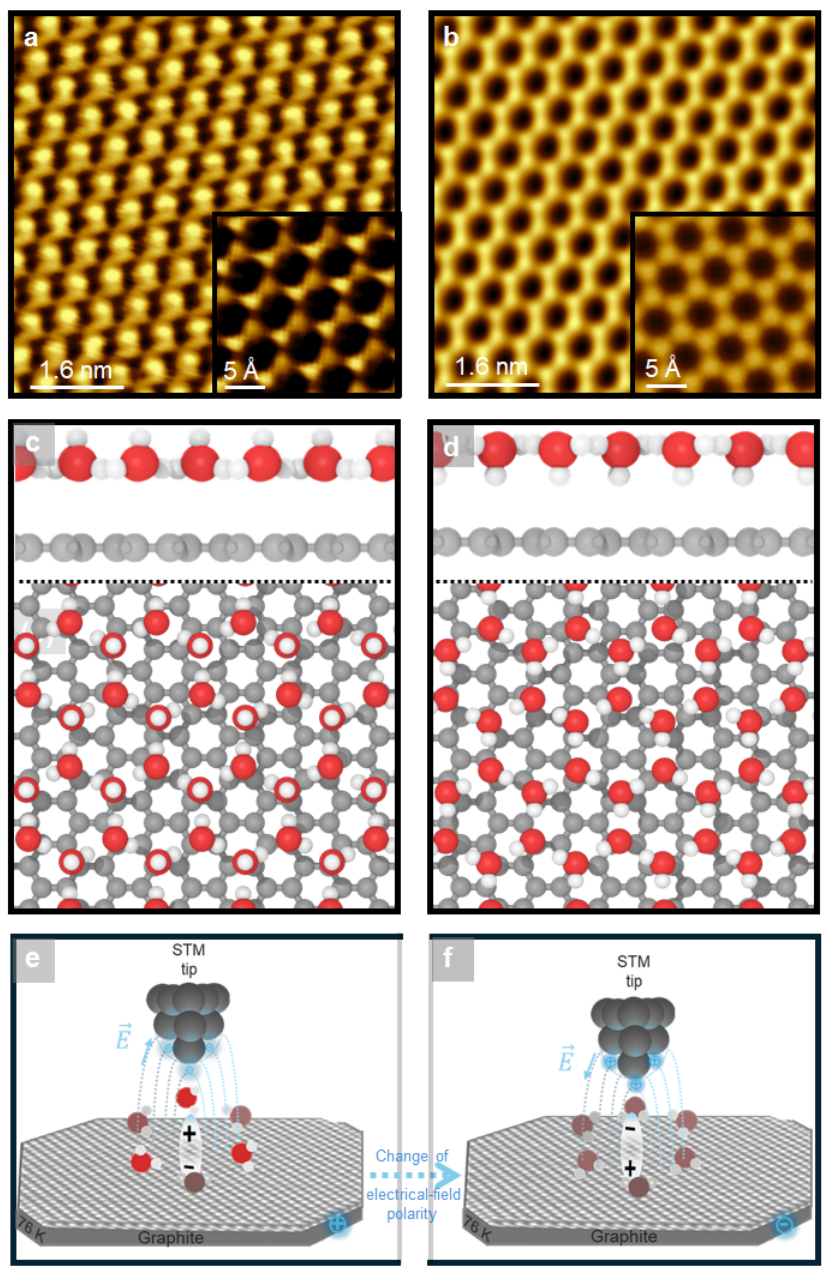

**Fig. 4 | Electric-field polarity-driven switching of a monolayer ice lattice on graphite.
a,** Empty-state STM image of the ice monolayer (8 × 8 nm$^2$, $V_s$ = +0.6 V, $I_t$ = 300 pA) acquired at 76 K. Inset: high-resolution image (2.5 × 2.5 nm$^2$, $V_s$ = +0.55 V, $I_t$ = 300 pA). **b,** Filled-state STM image of the same region (8 × 8 nm$^2$, $V_s$ = +0.6 V, $I_t$ = 300 pA), showing contrast inversion while preserving lattice symmetry. Inset: high-resolution image (2.5 × 2.5 nm$^2$, $V_s$ = +0.55 V, $I_t$ = 300 pA). **c-d,** Optimized atomic structures of monolayer ice on graphene corresponding to opposite molecular dipole orientations (H-up and H-down). **e&f,** Schematic illustrations of electric-field-induced switching of the hydrogen-bond network. Molecular dipoles are represented by ± symbols, and cyan lines indicate the electric field direction within the STM junction.

# Supplementary Information

# Electric-field control of hydrogen bonding via interfacial charge at atomic resolution

Nassar Doudin[1†]*, Jian Jiang[2†], Chun Tang[3,4], Xiao Cheng Zeng[5,6]* and Mohammed Th. Hassan[1]*.

[1]Department of Physics, University of Arizona, Tucson, AZ 85721, USA.

[2] Shenzhen Research Institute, City University of Hong Kong, Shenzhen, P. R. China.

[3]Faculty of Civil Engineering and Mechanics, Jiangsu University, Zhenjiang, Jiangsu, China.

[4] Suzhou Laboratory, Suzhou, China.

[5] Department of Material Science and Engineering, City University of Hongkong, Kowloon, Hongkong, China.

[6]Hong Kong Institute for Clean Energy, City University of Hong Kong, Kowloon, Hong Kong, China

[†] These authors contributed equally: Nassar Doudin, Jian Jiang

* Corresponding authors: ndoudin@arizona.edu, xzeng26@cityu.edu.hk and mohammedhassan@arizona.edu

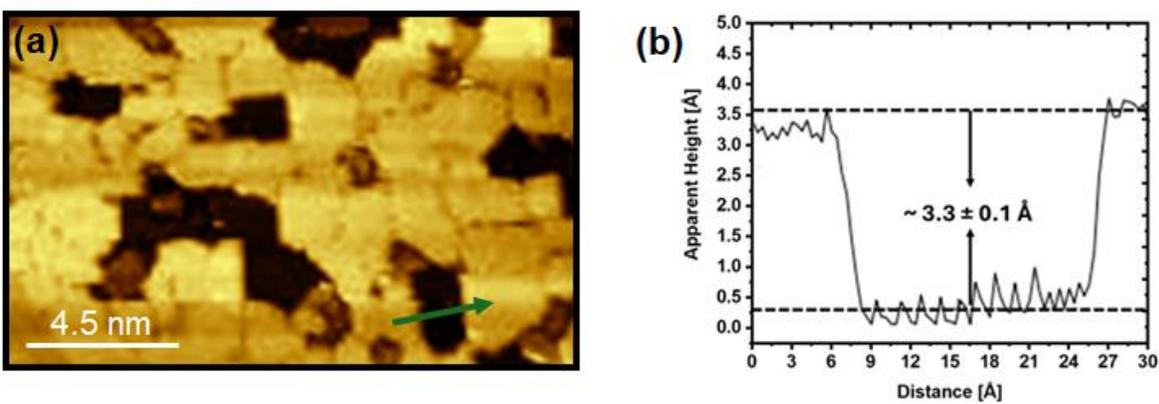

**Fig. S1 | Submonolayer ice coverage and height characterization on graphite.**
**a,** STM image of water adsorbed on graphite at 76 K, corresponding to a submonolayer coverage of ~0.7 ML (23 × 14 nm$^2$, $V_S$ = - 0.90 V, $I_t$ = 200 pA). **b,** Line profile extracted along the green arrow in a, showing an apparent depth of ~3.3 Å associated with a pit-like feature in the ice layer. To quantitatively establish the structural identity of the ice layer in the submonolayer regime, we examine water adsorption on graphite at an intermediate coverage of ~0.7 ML. Under these conditions, STM imaging reveals partial surface coverage, with well-defined ice domains coexisting with exposed graphite (Fig. S2a). This configuration enables a direct measurement of the height difference between the adsorbed layer and the substrate within the same field of view. A line profile acquired across the ice–graphite interface yields an apparent height difference of ~3.3 Å (Fig. S2b). This value is consistent with the expected thickness of a single ice layer on graphite and is significantly smaller than that of multilayer ice, thereby excluding the presence of thicker water aggregates.

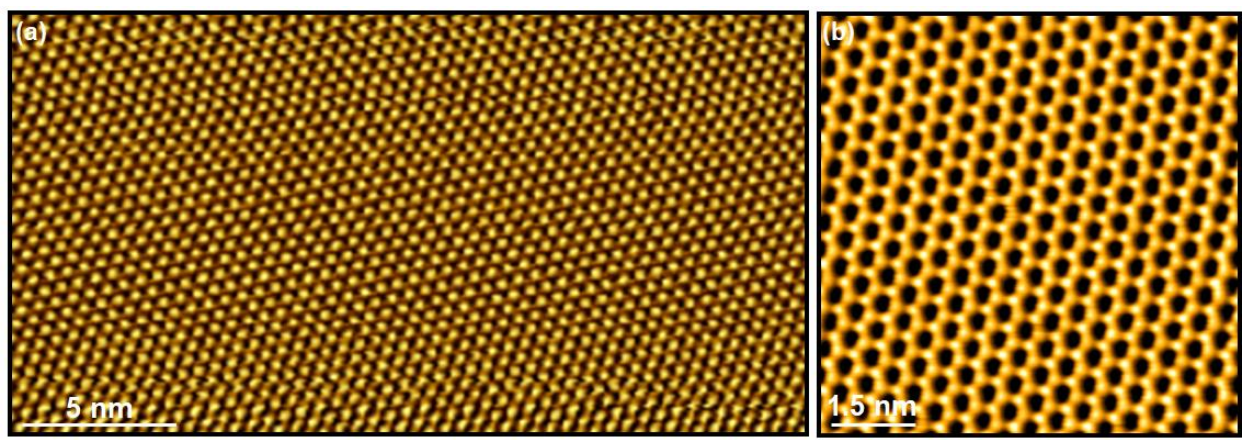

**Fig. S2 | Formation of a saturated monolayer ice film on graphite. a,b,** STM images of the graphite surface following deposition of a saturation dose of water at 76 K, showing the formation of a continuous, defect-free single-phase ice monolayer. Image conditions: **a,** (25 × 13 nm$^2$, $V_S$ = 0.73 V, $I_t$ = 300 pA); **b,** (7.5 × 7.5 nm$^2$, $V_S$ = 0.58 V, $I_t$ = 300 pA). At higher spatial resolution, the film exhibits a homogeneous morphology over tens of nanometres, with no detectable discontinuities, domain boundaries or phase coexistence (Fig. S1a). The uniform contrast across the terrace further indicates that the ice layer forms a single, laterally continuous two-dimensional phase. Notably, the ring structures (Fig. 1b) remain intact and undistorted, confirming the structural uniformity of the hydrogen-bond network.

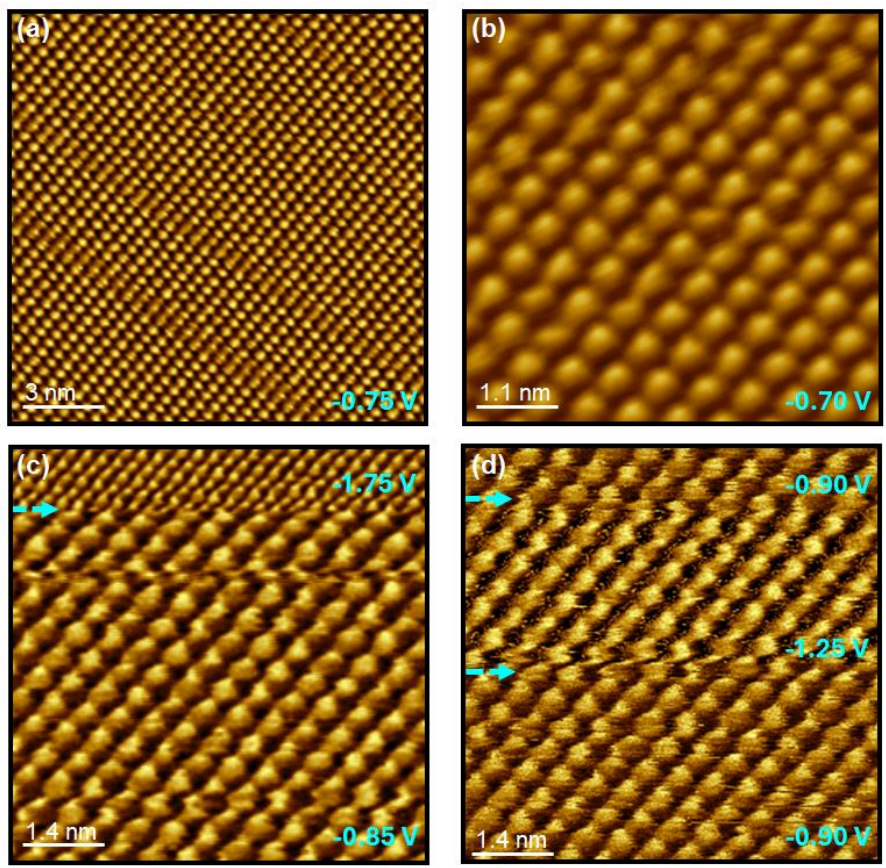

**Fig. S3 | Bias-dependent STM imaging of monolayer ice on graphite at negative bias. a–d,** STM images acquired at 76 K under negative sample bias, illustrating the dependence of the apparent structure on the applied bias voltage ($I_t$ = 250 pA; bias voltages and image sizes as indicated). The images show the evolution of the hydrogen-bond network with increasing electric field magnitude. The cyan arrow marks the boundary between regions acquired at different bias voltages within the same scan. Figure S3 presents a bias-dependent STM image sequence acquired at negative voltage bias, analogous to Fig. 2a–e (positive bias), demonstrating that the electric-field-driven evolution of the hydrogen-bond network is independent of bias polarity. As the magnitude of the negative bias is increased, the monolayer ice exhibits the same sequence of regimes observed at positive bias (Fig. 2): a well-ordered hexagonal structure at moderate bias, followed by a dynamically reconfigurable state with molecular-scale fluctuations, and ultimately destabilization leading to melting and desorption at higher biases. The reproducibility of these transitions under opposite polarity confirms that the observed behaviour is governed by the electric field rather than the sign of the applied voltage. These results establish that the field symmetrically reshapes the energy landscape of the hydrogen-bond network, enabling deterministic control of its structural and electronic states irrespective of polarity.

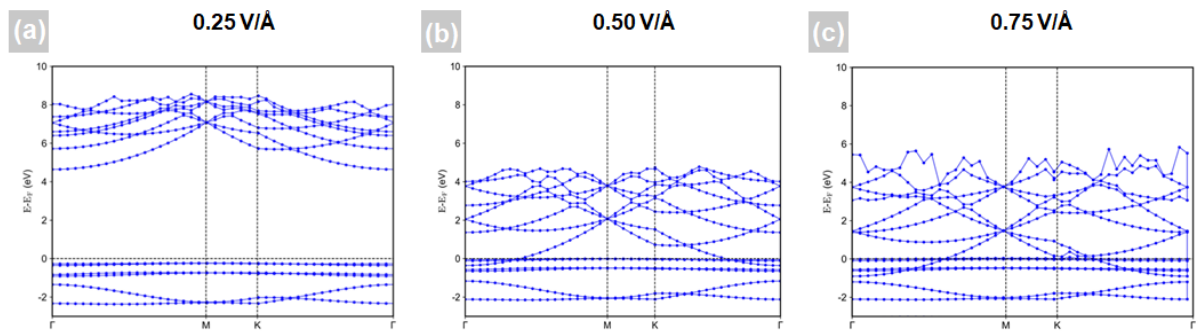

**Fig. S4 |** Band structure of monolayer ice at electric fields of 0.25, 0.5 and 0.75 V Å⁻¹.

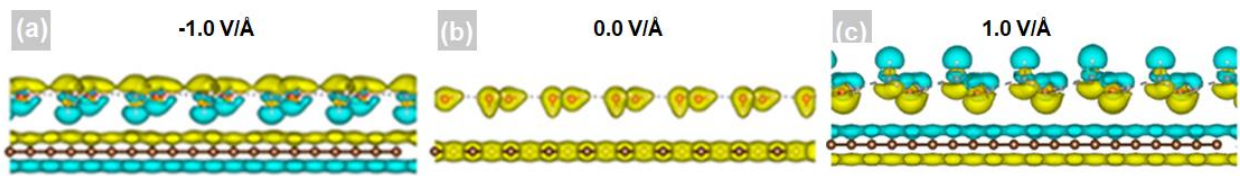

**Fig. S5 |** Charge difference distribution relative to the zero-field system, obtained from DFT calculations at electric fields of 1, 2 and 3 V Å⁻¹.